\documentclass[aps,prd,twocolumn,amsmath,amssymb,amsfonts,nofootinbib,superscriptaddress,altaffilletter]{revtex4-1}

\usepackage{graphicx}
\usepackage{hyperref}
\hypersetup{
  bookmarksopen=true
}
\usepackage{amssymb}
\usepackage{amsmath}
\usepackage{color}
\usepackage{supertabular}
\usepackage{dcolumn}

\def\Ap{A_+}
\def\Ax{A_{\times}}
\def\hp{h_+}
\def\hx{h_{\times}}
\def\Fp{F_+}
\def\Fx{F_{\times}}
\def\CSp{\textrm{CS}_+}
\def\CSx{\textrm{CS}_{\times}}
\def\Var{\textrm{Var\,}}
\def\expect{\mathop{{}\mathbb{E}}}

\def\sci#1#2{#1\times10^{#2}}
\def\widebar{\overline}

\def\etal{{\it et al.}}

\def\fc{\frac{1}{\sqrt{2\pi}}}

\begin{document}

\title{Loosely coherent searches for medium scale coherence lengths}

\author{Vladimir Dergachev}
\email{vladimir.dergachev@aei.mpg.de}
\affiliation{Max Planck Institute for Gravitational Physics (Albert Einstein Institute), Callinstrasse 38, 30167 Hannover, Germany}
\affiliation{Leibniz Universit\"at Hannover, D-30167 Hannover, Germany}

\begin{abstract}
The search for continuous gravitational waves demands computationally efficient algorithms that can handle highly non-linear parameter spaces. 
Loosely coherent algorithms establish upper limits and detect signals by analyzing families of templates as a single unit. We describe a new computationally efficient loosely coherent search intended for all-sky searches over medium scale coherence lengths ($3$\,--\,$300$\,hours).
\end{abstract}

\maketitle

\section{Introduction}

In this paper we describe a novel {\em Loosely coherent} algorithm for detecting continuous almost monochromatic signals. Such signals are searched for in data collected by LIGO and Virgo gravitational wave interferometers. The data under analysis typically spans several months and searches \cite{O1EH, O1AllSky1, O1AllSky2} look for signals in the $20$--$2000$\,Hz frequency band.

Carrying out an all-sky search in this data typically requires several months of computation on large clusters. Large numbers of outliers are produced in initial stages due to imperfect data and must be dealt with in subsequent analysis. A successful search must combine efficient algorithms with top-notch technical implementation and excellent search execution.

In this paper we focus on the mathematical aspects of the newly implemented {\em Loosely coherent} code. Such description eases reuse of the algorithm in other searches, and can also serve as a reference for results papers using the new algorithm. We present validation results of a search pipeline built using this algorithm. A speed comparison to previous implementation shows more than 10x speedup.

The {\em Loosely coherent} algorithms were introduced in \cite{loosely_coherent}, as a general approach to signal detection based on exploring regions of parameter space at once. This contrasts with methods that focus on detecting a single template \cite{jks} and then iterate over template banks. These methods were first applied to outlier followup and to searches of small sky areas. The new {\em loosely coherent} algorithm described in this paper has been developed to carry out all-sky searches and to deal with scalability issues discovered in applications of early {\em loosely coherent} methods.

The algorithm implementing optimal statistic depends strongly on parameter space region morphology. For example, if the family of potential signals is only described by occupied bandwidth then the optimal statistic is a filter extracting this bandwidth followed by a power detector \cite{loosely_coherent, zubakov}. On the other hand, long term drifts in the bandwidth window can be accommodated by dynamic programming algorithms which are fundamentally non-linear \cite{viterbimethod, viterbi, O1AllSky2}. 

Some parameter spaces allow themselves to be split into regions for independent analysis without compromising efficiency. Such regions can be labeled by a representative signal waveform which can then be considered as a ``thick'' template. This situation occurs in all-sky analysis where signal modulations are bounded, and usually no larger than  $10^{-4}\cdot f$ at high signal frequencies $f$. 

For example, one can split the bandwidth into $1/8$\,Hz bands which, effectively, creates a thick template bank. In practice, the partition is usually much finer and is done in all the search parameters. This is necessary to reduce memory footprint of practical implementation.

The signal families describing potential continuous gravitational wave signals generated by rotating neutron stars are complex. Even in the case of an isolated neutron star the incoming gravitational wave is expected to slowly drift in frequency, which is usually approximated with a linear term and sometimes quadratic term. By the time it reaches the detector the signal undergoes strong Doppler shifts from complex motion\footnote{Exact computation of phase corrections requires transcendental functions, see \cite{barycenter} for a computationally efficient approximation by polynomials} of the detector around Earth rotation axis and along Earth orbit around the Sun Relativistic effects come into play and the signals can be delayed due to gravitational field of Sun or other bodies \cite{barycenter}.

The first implementation \cite{loosely_coherent, orionspur} of a {\em Loosely coherent algorithm} computed power $P[f]$ as a function of frequency $f$. The power function $P[f]$ is quadratic sum of input data $\left\{a_{t,f}\right\}$:
\begin{equation}
\label{eqn:power_sum}
P[f] = \sum_{t_1, t_2} a_{t_1, f+\delta f(t_1)} a_{t_2, f+\delta f(t_2)}^* K_{t_1, t_2, f}
\end{equation}
Here $t_1$ and $t_2$ denote times of frequency bins $\left\{a_{t,f}\right\}$ of short Fourier transforms (SFTs) of the data acquired by the detectors, $f$ is the frequency of the desired signal, $\delta f(t)$ is the frequency shift of the signal received by the detectors.

This initial implementation focused on kernels $K_{t_1, t_2, f}$ with non-zero entries in a narrow diagonal band $|t_1-t_2|<T$, where $T$ depended on effective coherence length (usually a constant fraction of $T$) of the search. The input data was phase-corrected so that transformed data approximated a heterodyned signal that would have been received in Solar System Barycenter (SSB) frame\cite{loosely_coherent}. The Lanzcos kernel was used that combined a low-pass filter and a power detector. 

This approach made sense because that {\em Loosely coherent} algorithm was used for followup of outliers produced by an algorithm that computed power in individual SFT bins and required no phase corrections. The diagonal band was initially very short and then increased in subsequent stages. 

%Partitioning of computation by Doppler shifts and an associative cache were used to increase speed.

However, as the width of diagonal band (and effective coherence length) was increased the speedup techniques would eventually become ineffective and the computational demands would scale quadratically with coherence length even for a single template, due to summation over two indices $t_1, t_2$. Thus the algorithm was only practical  for short coherence lengths (of a small multiple of SFT length) or when few templates would be searched compared to previous stages, such as during followup of outliers. 

Long coherence lengths were explored with {\em Loosely coherent} algorithm designed for well-modeled signals \cite{loosely_coherent2}, which scaled to coherence lengths of more than $10^6$\,s. While longer coherence length generally results in more sensitivity, the improvement in signal-to-noise ratio (SNR) scales only as forth root of increase in coherence length, while the size of parameter space scales as a forth power or faster. 

Even though {\em Loosely coherent} algorithm for well-modeled signals was spending less than 1500 CPU cycles per template (amortized over the entire computation) \cite{loosely_coherent2} the forth power scaling made searches of large portions of the sky impractical on contemporary computer clusters. 

This paper describes how to construct {\em Loosely coherent} algorithm for medium scale coherence lengths of $3$\,--\,$300$\,hours, assuming input data of 3600\,s short Fourier transforms (SFTs). The new algorithm is practical for all-sky searches using several hours long coherence time on contemporary computer clusters.

\section{Loosely coherent sum}
In order to construct a practical algorithm we must find an efficient way to compute the power sum in equation \ref{eqn:power_sum}.

For long coherence length the power sum kernel $K_{t_1, t_2, f}$ must have a lot of non-zero terms making straightforward summation impractical. A well-known way \cite{zubakov} to speed up computation is to decompose total power into the sum of squares of coherent combinations of phase-corrected terms:
\begin{equation}
P[f] = \sum_{t_1} \left|CS[t_1, f] \right|^2
\end{equation}
with coherent sums $CS$ given by
\begin{equation}
CS[t_1, f]= \sum_{t_2} w_{t_1t_2} e^{i\phi(t_2)} a_{t_2, f+\delta f(t_2)}
\end{equation}
The effective coherence length is then governed by the number of terms in the coherent sum $CS$ and the computational load will scale linearly at this stage of the computation. 

Both weights $w_{t_1t_2}$ and phase corrections $\phi(t_2)$ vary over parameter space, however variation of weights is very gradual, so it makes sense to design our statistic assuming weights can be constant and precomputed over extended area. We will discuss weights in more detail in the subsequent section.

Because we are only interested in the absolute value of the coherent sum the variation of phase corrections over parameter space reduces to a second order difference operator:
\begin{equation}
\phi(t, p_1)-\phi(t, p_2) \sim \phi(t, p_1)-\phi(t_0, p_1) -  \phi(t, p_2)+\phi(t_0, p_2)
\end{equation}
Here $p_1$ and $p_2$ are two locations in the parameter space, $t$ is some point of time for which we are interested in a phase correction, and $t_0$ is some fixed reference time, such as a midpoint of data being used in a single coherent sum. Because we are interested in power, the constant phase correction $\phi(t_0, p_1)-\phi(t_0, p_2)$ can be neglected.

The second order difference operator removes large portion of nonlinearity from Barycentric corrections, with the remainder described as relatively simple algebraic expression in functions of time and sky location \cite{barycenter}.

Thus the differences are predictable based on local data. One straightforward way to take advantage of this is with an associative cache of previously computed coherent sums - if the expected phase differences are close enough we can reuse already computed result.

A more sophisticated approach being considered for future implementation is to use the explicit algebraic form of the differences \cite{barycenter} to compute power sums over a wider area on the sky. The algebraic form (or, at least, knowledge of the Lipschitz constant) is also essential for application of cubic algorithm \cite{cubic}, which performs global optimization.

\section{Coherent sum weights}

We now address the question of how to pick weights when the noise level of underlying data $a_{t,f}$ varies. Because the weights in the different coherent sums can (and should be) chosen independently we will focus on a single coherent sum at a time.

To simplify notation we focus on a specific frequency $f$ and introduce coherent sum elements $z_i$ which are rescaled from $a_{t,f}$ so that the signal contributes the same amplitude and phase $Ae^{i\phi}$ to each $z_i$. The index $i$ runs over individual SFTs used in the coherent sum. Because amplitude modulation needs to be backed out (discussed in the following section) the variance of noise component of $z_i$ is expected to not be stationary.

We thus estimate power by computing absolute squared value of the weighted coherent sum $CS$
\begin{equation}
\textrm{P}=P_0+\left|CS\right|^2=P_0+\left|\sum_{i=1}^N  w_i z_i\right|^2
\end{equation}
where $P_0$ denotes contribution of other coherent sums, weights $w_i$ are real and non-negative, while the coherent sum elements $z_i$ are usually complex numbers. Normalization requires that the weights $w_i$ should sum to $1$.

Let us assume that the noise in $z_i$ is uncorrelated. We will now find the optimal set of weights $w_i$ that minimizes $U=\Var \left|CS\right|^2$. 
This choice of the utility function is somewhat arbitrary, but the variance works well for common distributions like Gaussian or exponential and is easy to compute. It is not appropriate for quantifying spiky noise - this is best analyzed with other data quality techniques. In our case, we are interested in the behavior of the middle portion of the distribution.

Also, ideally, we should optimize the variance of the entire power sum $P$, but this would introduce dependence between weights of separate coherent sums when the coherent sums use overlapping stretches of data. In case of no overlap  the coherent sums can be assumed independent and optimizing $\Var P$ is the same as separately optimizing variance of each coherent sum.

We compute
\begin{equation}
\begin{array}{l}
 U=\Var \left|CS\right|^2=\Var\left(\sum\limits_{i,j=1}^N w_iw_j z_i \bar{z}_j\right)=\\
% \quad\quad=\sum\limits_{i=1}^N w_i^4 \Var |z_i|^2 +\\
% \quad\quad\quad\quad+4\sum\limits_{1=i<j}^N w_i^2w_j^2 \left(\expect (\Re(z_i\bar{z}_j))^2-\left(\expect \Re(z_i\bar{z}_j)\right)^2\right)=\\ 
% \quad\quad=\sum\limits_{i=1}^N w_i^4 \Var |z_i|^2 +\\
% \quad\quad\quad\quad+4\sum\limits_{1=i<j}^N w_i^2w_j^2 \left(\expect (\Re z_i \Re z_j +\Im z_i \Im z_j)^2-\left(\Re \left( \expect z_i \expect \bar{z}_j\right)\right)^2\right)=\\ 
 \quad=\sum\limits_{i=1}^N w_i^4 \Var |z_i|^2 +\\
 \quad\quad+4\sum\limits_{1=i<j}^N w_i^2w_j^2 \left(\expect (\Re z_i)^2 \expect ( \Re z_j)^2 +\expect(\Im z_i)^2\expect( \Im z_j)^2+\right.\\
 \quad\quad\left.+2\expect(\Re z_i \Im z_i)\expect(\Re z_j \Im z_j)-\left(\Re \left( \expect z_i \expect \bar{z}_j\right)\right)^2\right)\\ 
\end{array}
 \end{equation}

Let us consider the situation of background noise with no signal. In this case the phases would be randomly distributed and we can assume $\expect \Re z_i=\expect \Im z_i=\expect \Re z_i \Im z_i=0$. Also randomness of phases implies $\expect (\Re z_i)^2=\expect (\Im z_i)^2=\expect |z_i|^2/2$. Thus
\begin{equation}
\begin{array}{l}
 U_{\textrm{bg}}=\sum\limits_{i=1}^N w_i^4 \Var |z_i|^2+2\sum\limits_{1=i<j}^N w_i^2w_j^2\expect |z_i|^2 \expect |z_j|^2
\end{array}
\end{equation}

If $z_i$ are Gaussian then $|z_i|^2$ is exponentially distributed and $\Var |z_i|^2=\left(\expect |z_i|^2\right)^2$. Thus
\begin{equation}
\begin{array}{l}
 U_{\textrm{Gauss bg}}=\sum\limits_{i=1}^N w_i^4 \left(\expect |z_i|^2\right)^2+2\sum\limits_{1=i<j}^N w_i^2w_j^2\expect |z_i|^2 \expect |z_j|^2=\\
 \quad\quad=\left(\sum\limits_{i=1}^N w_i^2 \expect |z_i|^2 \right)^2
\end{array}
\end{equation}

Thus for a Gaussian $z_i$ with randomly distributed phases minimizing $U$ is equivalent to minimizing $\sum\limits_{i=1}^N w_i^2 \expect |z_i|^2$. The optimum is easily found:
\begin{equation}
w_i=\frac{1}{\mathcal{N}}\frac{1}{\expect |z_i|^2}=\frac{1}{\mathcal{N}}\frac{1}{\sqrt{\Var |z_i|^2}} 
\end{equation}
Here ${\mathcal{N}}$ is a suitable normalizing factor.

\section{Polarization analysis}
We will now derive explicit formulas for loosely coherent power sums.

Let us assume that the incoming gravitational wave signal is a monochromatic wave represented as \cite{jks}
\begin{equation}
\begin{array}{l}
\hp=\Ap \cos(\omega t+\phi) \cos(\epsilon)-\Ax \sin(\omega t+\phi) \sin(\epsilon) \\
\hx=\Ap \cos(\omega t+\phi) \sin(\epsilon)+\Ax \sin(\omega t+\phi) \cos(\epsilon)
\end{array}
\end{equation}
where $\epsilon=2\psi$ describes orientation of the pulsar, $\phi$ is the initial signal phase, $\omega=2\pi f$ is the pulsar gravitational wave frequency and $\Ap=h_0(1+\cos^2(\iota))/2$ and $\Ax=h_0\cos(\iota)$ are amplitudes. 

The signal received by the detector is described by
\begin{equation}
s(t)=\Fp(t)\hp +\Fx(t) \hx 
\end{equation}
where $\Fp(t)$ and $\Fx(t)$ are time-varying amplitude response variables of the detector.

The input data consists of short Fourier transforms with coherence length small enough that one can assume that $\Fp(t)$ and $\Fx(t)$ do not vary significantly.

The frequency bin $a_{t,f}$ of the SFT taken at time $t$ can thus be described as
\begin{equation}
\begin{array}{l}
a_{t,f}=n_{t,f}+\Fp(t)(\Ap e^{i\Phi(t)} \cos(\epsilon)+\Ax i e^{i\Phi(t)} \sin(\epsilon)) +\\
\quad\quad\quad\quad+\Fx(t)(\Ap e^{i\Phi(t)} \sin(\epsilon)-\Ax i e^{i\Phi(t)} \cos(\epsilon))=\\
%\quad\quad=n_{t,f}+h_0e^{i\Phi(t)} \left(\Fp(t)(\frac{1+\cos^2(\iota)}{2}  \cos(\epsilon)+i\cos(\iota) \sin(\epsilon)) +\right.\\
%\quad\quad\quad\quad+\left.\Fx(t)(\frac{1+\cos^2(\iota)}{2}\sin(\epsilon)- i\cos(\iota) \cos(\epsilon))\right)\\
\quad\quad=n_{t,f}+h_0e^{i\Phi(t)} \left(\Fp(t)w'_1 +\Fx(t)w'_2\right)\\
\end{array}
\end{equation}
where $n_{t,f}$ is the detector noise. The variables $w'_1$ and $w'_2$ are complex amplitude parameters \cite{loosely_coherent2}, they satisfy
\begin{equation}
\sqrt{|w'_1+iw'_2|}+\sqrt{|w'_1-iw'_2|}=1 
\end{equation}

The coherent power sum computed over SFTs taken at times $t_1,\ldots t_N$ is
\begin{equation}
\begin{array}{l}
\textrm{PS}=\left|\sum_{i=1}^N  w_i a_{t_i,f_i}e^{-i\Phi(t_i)}/\left(\Fp(t_i)w'_1+\Fx(t_i)w'_2\right)  \right|^2
\end{array}
\end{equation}
where we used
\begin{equation}
z_i=a_{t_i,f_i}e^{-i\Phi(t_i)}/\left(\Fp(t_i)w'_1+\Fx(t_i)w'_2\right)
\end{equation}

The division by amplitude response of the detector effectively increases noise level of SFTs at time of unfavorable orientation. Thus we expect the optimal weights to deweight these terms.

Assuming $n_{t_i, f_i}$ are Gaussian the optimal weights computed in the previous section are:
\begin{equation}
w_i=\left|\Fp(t_i)w'_1+\Fx(t_i)w'_2\right|^2/\expect \left|n_{t_i,f_i}\right|^2
\end{equation}
The normalizing factor is
\begin{equation}
\mathcal{N}=\sum_{i=1}^n w_i
\end{equation}

And thus the optimal power sum is
\begin{equation}
\textrm{PS}=\frac{1}{\mathcal{N}^2}\left|\sum_{i=1}^N  a_{t_i,f_i}e^{-i\Phi(t_i)} \frac{\Fp(t_i)\bar{w}'_1+\Fx(t_i)\bar{w}'_2}{\expect \left|n_{t_i, f_i}\right|^2}  \right|^2
\end{equation}

The normalizing factor can be expressed as
\begin{equation}
\begin{array}{l}
\mathcal{N}=\left(\sum_{i=1}^N \Fp(t_i)^2/\expect \left|n_{t_i,f_i}\right|^2\right) \left|w'_1\right|^2+\\
\quad\quad\quad\quad+\left(\sum_{i=1}^N \Fx(t_i)^2/\expect \left|n_{t_i,f_i}\right|^2\right) \left|w'_2\right|^2+\\
\quad\quad\quad\quad+2\left(\sum_{i=1}^N \Fx(t_i)\Fp(t_i)/\expect \left|n_{t_i,f_i}\right|^2\right) \Re\left(w'_1 \widebar{w'_2}\right)\\
\end{array}
\end{equation}

Let us define two intermediate coherent sums $\CSp$ and $\CSx$:
\begin{equation}
 \begin{array}{l}
 \CSp=\sum_{i=1}^N  a_{t_i,f_i}e^{-i\Phi(t_i)}\Fp(t_i)/\expect \left|n_{t_i, f_i}\right|^2  \\
 \CSx=\sum_{i=1}^N  a_{t_i,f_i}e^{-i\Phi(t_i)}\Fx(t_i)/\expect \left|n_{t_i, f_i}\right|^2 
 \end{array}
\end{equation}

Then the power sum is expressed as
\begin{equation}
 \begin{array}{l}
\textrm{PS}=\frac{1}{\mathcal{N}^2}\left(|\CSp|^2 |w'_1|^2+|\CSx|^2 |w'_2|^2+\right. \\
\quad\quad\quad\quad\quad\quad\quad\quad\quad \left.+ 2\Re\left(\CSp \widebar{\textrm{CS}}_{\times} w'_2 \bar{w}'_1\right)\right)
 \end{array}
\end{equation}
or
\begin{equation}
 \begin{array}{l}
\textrm{PS}=\frac{1}{\mathcal{N}^2}\left(|\CSp|^2 |w'_1|^2+|\CSx|^2 |w'_2|^2+\right.\\
\quad\left.+2\Re\left(\CSp \widebar{\textrm{CS}}_{\times}\right)\Re\left(w'_1 \bar{w}'_2\right)+2\Im\left(\CSp \widebar{\textrm{CS}}_{\times}\right)\Im\left(w'_1 \bar{w}'_2\right)\right)
 \end{array}
\end{equation}
We see that the power sum is a rational function of coefficients $w'_1$ and $w'_2$. This allows to postpone substitution of these coefficients until after summation is performed. In a typical search the parameters $w'_1$ and $w'_2$ should be sampled on a grid containing hundreds of points in order to minimize power loss, thus there is great advantage that only 7 separate sums (4 for the numerator and 3 for the normalizing factor $\mathcal{N}$) need to be accumulated before substitution.

For convenience we list some common expressions of $w'_1$ and $w'_2$ in terms of more conventional parameters $\iota$ and $\epsilon$ \cite{S4IncoherentPaper, EarlyS5Paper, FullS5Semicoherent, S6PowerFlux, O1AllSky2}:
\begin{equation}
 \begin{array}{rcl}
w'_1&=& \frac{1+\cos^2(\iota)}{2}  \cos(\epsilon)+i\cos(\iota) \sin(\epsilon)\\
w'_2&=& \frac{1+\cos^2(\iota)}{2}\sin(\epsilon)- i\cos(\iota) \cos(\epsilon) \\
|w'_1|^2&=& \frac{1+2\cos^2(\iota)+\cos^4(\iota)}{4} \cos^2(\epsilon)+\cos^2(\iota) \sin^2(\epsilon)  \\
|w'_2|^2&=& \frac{1+2\cos^2(\iota)+\cos^4(\iota)}{4} \sin^2(\epsilon)+\cos^2(\iota) \cos^2(\epsilon) \\
\Re\left(w'_1 \widebar{w'_2}\right)&=&\frac{\sin^4(\iota)}{4} \sin(\epsilon)\cos(\epsilon) \\
\Im\left(w'_1 \widebar{w'_2}\right)&=& \frac{1+\cos^2(\iota)}{2}\cos(\iota)\\
 \end{array}
\end{equation}

\section{Estimation of input data}
So far we have assumed that the source signal is bin-centered in the input SFT data $a_{t,f}$. In practice, this is usually not so. Thus the ideal, bin-centered value needs to be estimated from discretely sampled frequency bins of input SFTs.

To limit contamination from exceedingly steep detector artifacts these SFTs are computed using Hann window which minimizes spectral leakage. We thus need to compute which frequency bins of Hann-windowed SFTs contain putative signal and then devise a linear filter that would estimate signal amplitude in a fractional bin. 

In effect, the linear filter thus adds weighted summation of neighboring frequency bins to  the quadratic power sum $PS$. In practice, it is more efficient to compute the estimated fractional frequency bins on discrete grid and then look them up as needed.

\subsection{Hann response function}
Consider the function
\begin{equation}
X(t)=Ae^{2\pi ift}
\end{equation}

We compute Hann windowed frequency spectrum in bins $\frac{n}{t_1-t_0}$:
\begin{equation}
\begin{array}{l}
\label{hann_response}
F_n=\fc\frac{1}{t_1-t_0}{\displaystyle \int\displaylimits_{t_0}^{t_1} X(t)\left(1-\cos\left(2\pi \frac{t-t_0}{t_1-t_0}\right)\right) e^{- \frac{2\pi in(t-t_0)}{t_1-t_0}}dt}\\
%\quad\quad=\fc\frac{A}{t_1-t_0}\int_{t_0}^{t_1} e^{2\pi ift} \left(1-\cos\left(2\pi \frac{t-t_0}{t_1-t_0}\right)\right) e^{-2\pi i \frac{(t-t_0)n}{t_1-t_0}}dt\\
%\quad\quad=\fc A\int_{0}^{1} e^{2\pi ifs(t_1-t_0)+2\pi ift_0} \left(1-\cos\left(2\pi s\right)\right) e^{-2\pi i sn}ds\\
%\quad\quad=\fc Ae^{2\pi i ft_0}\int_{0}^{1} \left(1-\cos\left(2\pi s\right)\right) e^{2\pi is (f(t_1-t_0)-n)}ds\\
%\quad\quad=\fc Ae^{2\pi i ft_0} \frac{i}{2}\cdot\frac{-1+e^{2\pi i \delta}}{\pi \delta (\delta^2-1)}\\
%\quad\quad=\fc Ae^{2\pi i ft_0}e^{\pi i \delta}\cdot\frac{\sin(\pi \delta)}{\pi \delta (1-\delta^2)}\\
\quad\quad=\fc Ae^{2\pi i f (t_0+t_1)/2}e^{-\pi i n}\cdot\frac{\sin(\pi \delta)}{\pi \delta (1-\delta^2)}
\end{array}
\end{equation}

where $\delta=f(t_1-t_0)-n$.

This is an entire holomorphic function of $\delta$ which decays as $\delta^{-3}$ on the real axis.

\begin{figure}
  \begin{center}
  \includegraphics[height=8cm]{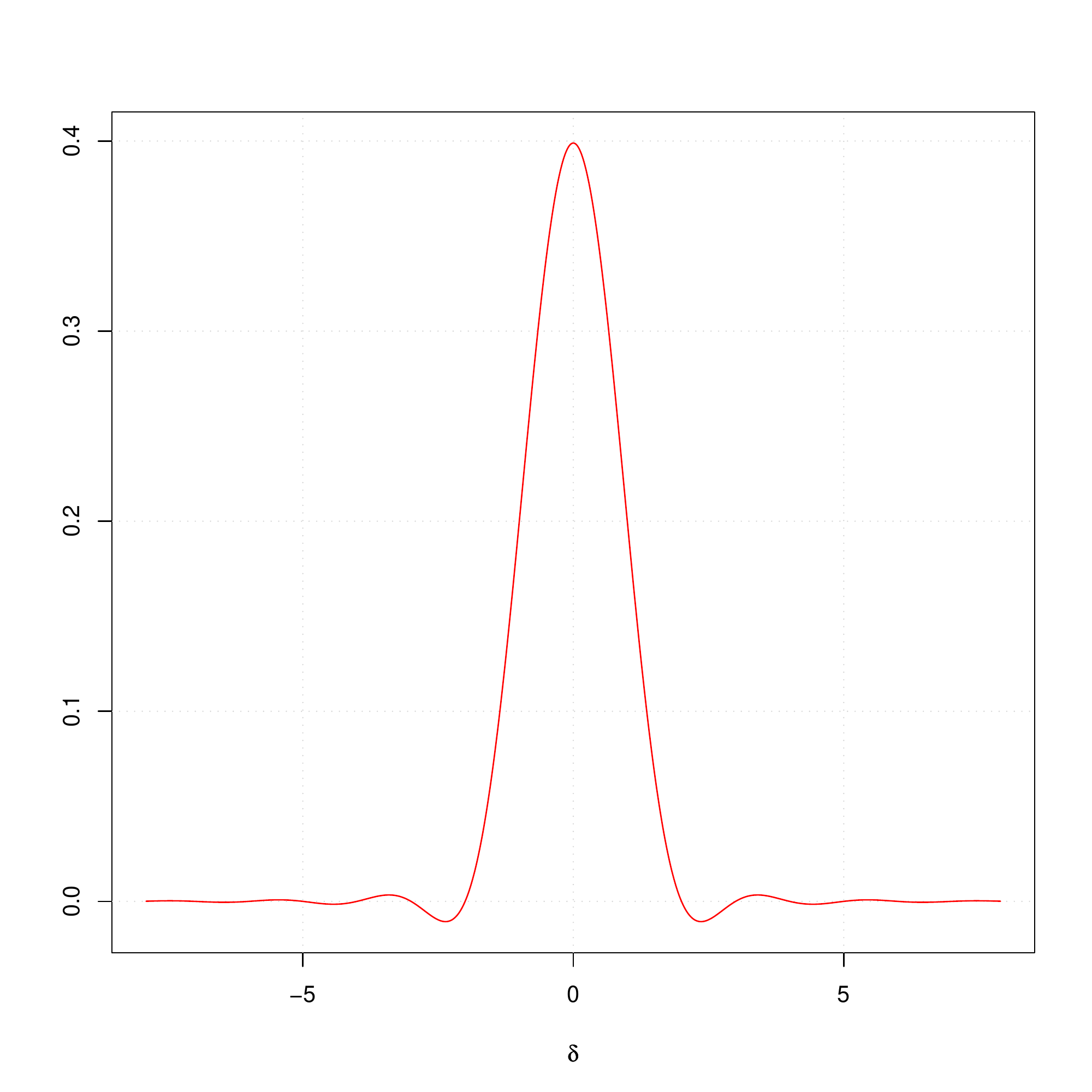}
  \caption{Amplitude of SFT bin for a test signal mismatched by $\delta$. This is a plot of $\fc \frac{\sin(\pi \delta)}{\pi \delta (1-\delta^2)}$, excluding exponential phase factors.}
  \label{fig:hann_response} 
  \end{center}
\end{figure}

\begin{figure}
  \begin{center}
  \includegraphics[height=8cm]{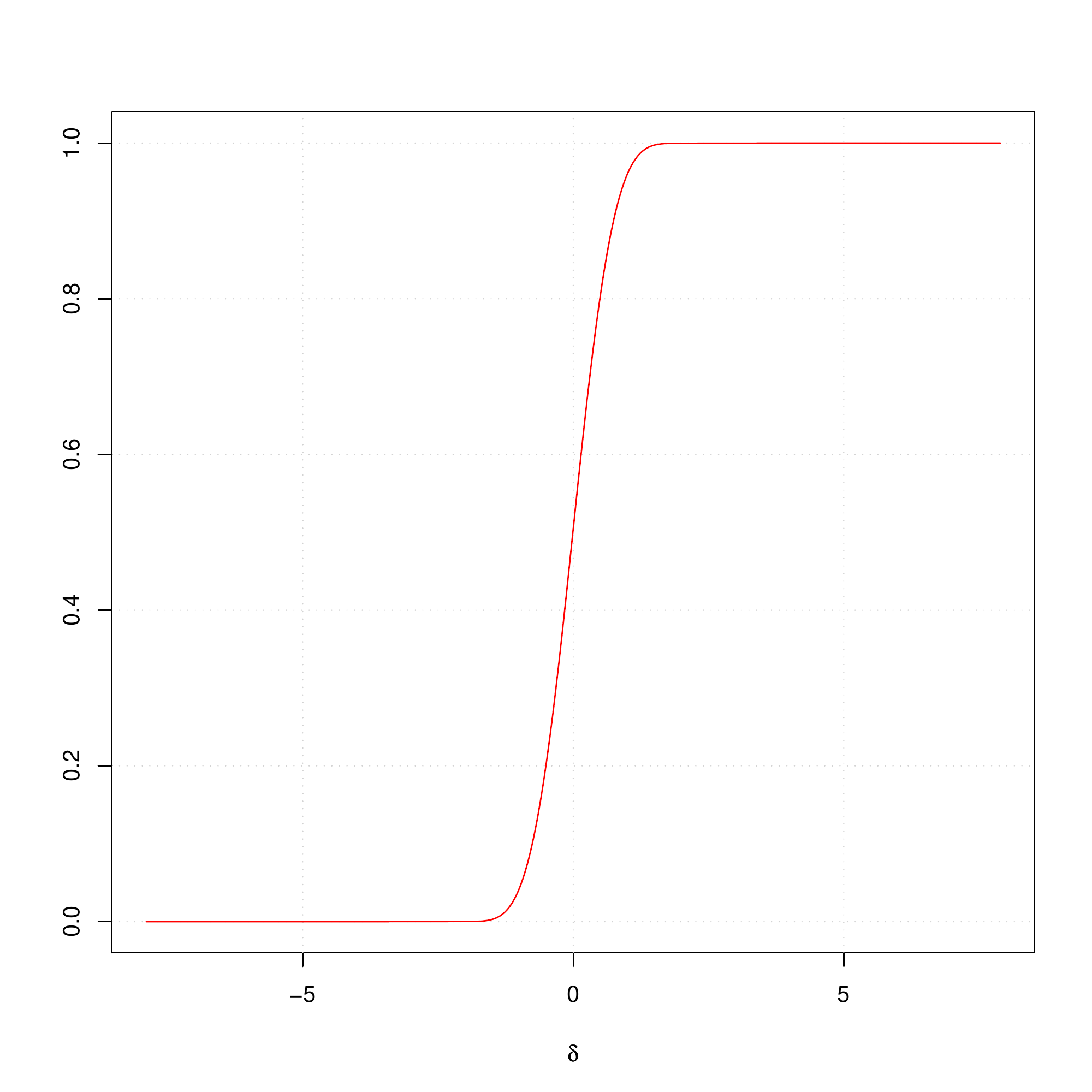}
  \caption{Integral of $|F_n|^2$, normalized to unity at $+\infty$. This shows total power in SFT bins below $f+\delta$ away from test signal frequency $f$.}
  \label{fig:hann_response_power} 
  \end{center}
\end{figure}

Figure \ref{fig:hann_response} shows the plot of $F_n$ without the exponential phase factors. One observes that for practical purposes the function vanishes for $|\delta|>3$.
Indeed, maximum over bins $n$ with $|\delta|>3$ divided by the value in the bin closest to the injected frequency (and thus with largest $F_n$) is below $0.01$. Power is even more tightly constrained with $99.97\%$ of it occurring in the interval $|\delta|<2$ (see figure \ref{fig:hann_response_power}).

\subsection{Matched filter estimation of signals at fractional frequency bins}
\label{matched_filter_description}

Suppose we have $M$ observations $y_i$ that are linear combinations of signal of unknown strength and noise:

\begin{equation}
y_i=Aa_i+\xi_i
\end{equation}
here $A$ is the (unknown) amplitude of our signal, $a_i$ are known and $\xi_i$ is the measurement noise.

The goal is to find linear combination $\tilde{A}=\sum_i \beta_i y_i$ with the best signal to noise ratio.

We have:
\begin{equation}
\tilde{A}=A \sum_i \beta_i a_i + \sum_i \beta_i \xi_i
\end{equation}

\begin{equation}
\textrm{SNR}=A \frac{\sum_i \beta_i a_i}{\sqrt{\sum_{i,j} \beta_i \beta_j \langle\xi_i , \xi_j\rangle}}
\end{equation}

Maximizing SNR is equivalent to minimizing $\sum_{i,j} \beta_i \beta_j \langle\xi_i , \xi_j\rangle$ while keeping 
$\sum_i \beta_i a_i$ constant.

Using method of Lagrange multipliers one easily obtains:
\begin{equation}
\sum_j  \langle\xi_i , \xi_j\rangle \beta_j = \lambda a_i
\end{equation}
where $\lambda$ is an arbitrary, but fixed constant.

Which leads to the well known result:
\begin{equation}
\vec{\beta}=\lambda R^{-1}\vec{a}
\end{equation}
where $R=\left|\langle\xi_i , \xi_j\rangle\right|_{i,j}$. It is usually convenient to pick $\lambda =  \left(\vec{a}^T R^{-1}\vec{a}\right)^{-1}$ - this ensures that a unit signal returns unit response.

For spectrum of Hann windowed signal 
\begin{equation}
\xi_i = -\frac{1}{2} \zeta_{i-1}+\zeta_i -\frac{1}{2} \zeta_{i+1}
\end{equation}
where $\zeta_i$ are independent identically distributed Gaussian variables with zero mean. Thus we can assume 
\begin{equation}
R_{ij}=\left\{\begin{array}{ll}
	\frac{3}{2} & i=j \\
	-1 & \left|i-j\right|=1 \\
	\frac{1}{4} & \left|i-j\right|=2 \\
	0 & \textrm{otherwise}
              \end{array}\right.
\end{equation}

It is instructive to compute $R$ and its inverse for the cases of $5$ and $7$ SFT bins. For $5$ sample points
\begin{equation}
R=\frac{1}{4}\left(\begin{array}{ccccccc}
6 & -4 & 1 & 0 & 0 &\\
-4 & 6 & -4 & 1 & 0 &\\
1 & -4 & 6 & -4 & 1 &\\
0 & 1 & -4 & 6 & -4 &\\
0 & 0 & 1 & -4 & 6 & \\
\end{array}\right)
\end{equation}

\begin{equation}
R^{-1}=\frac{1}{7}\left(\begin{array}{ccccc}
15 & 20 & 18 & 12 & 5 \\
20 & 40 & 40 & 28 & 12 \\
18 & 40 & 52 & 40 & 18 \\
12 & 28 & 40 & 40 & 20 \\
5 & 12 & 18 & 20 & 15
\end{array}\right)
\end{equation}

For $7$ sample points
\begin{equation}
R=\frac{1}{4}\left(\begin{array}{ccccccc}
6 & -4 & 1 & 0 & 0 & 0 & 0\\
-4 & 6 & -4 & 1 & 0 & 0 & 0\\
1 & -4 & 6 & -4 & 1 & 0 & 0\\
0 & 1 & -4 & 6 & -4 & 1 & 0\\
0 & 0 & 1 & -4 & 6 & -4 & 1\\
0 & 0 & 0 & 1 & -4 & 6 & -4\\
0 & 0 & 0 & 0 & 1 & -4 & 6\\
\end{array}\right)
\end{equation}

\begin{equation}
R^{-1}=\frac{1}{135}\left(\begin{array}{ccccccc}
336 & 504 & 540 & 480 & 360 & 216 & 84\\
504 & 1071 & 1260 & 1170 & 900 & 549 & 216\\
540 & 1260 & 1800 & 1800 & 1440 & 900 & 360\\
480 & 1170 & 1800 & 2100 & 1800 & 1170 & 480\\
360 & 900 & 1440 & 1800 & 1800 & 1260 & 540\\
216 & 549 & 900 & 1170 & 1260 & 1071 & 504\\
84 & 216 & 360 & 480 & 540 & 504 & 336
\end{array}\right)
\end{equation}

This procedure can be used for windows other than Hann, simply by updating matrices $R$. However, the number of terms and size of the matrices are minimized for Hann-windowed data.

The coefficients $a_i$ can be obtained from the formula \ref{hann_response}, however, if we are only interested in computing power $|\tilde{A}|^2$ we can discard the common phase factor depending on $f$ and simplify the formula to

\begin{equation}
a_k=(-1)^k\fc \cdot\frac{\sin(\pi \delta)}{\pi \delta (1-\delta^2)}
\end{equation}
where $\delta=f(t_1-t_0)-k$

\section{Dynamic programming methods}
Search over parameters which cause long-term frequency drift can be made more efficient by use of dynamic programming methods such as Viterbi algorithm \cite{viterbi, viterbimethod, O1AllSky2}.

In the simplest implementation, the input data is broken into $K$ chunks. Signal power for a set of frequency bins is computed in each chunk. The power is then accumulated progressively from first to last chunk and at each accumulation stage the power in a frequency bin is replaced by maximum power over a set nearby bins \cite{O1AllSky2}. 

For example, one can take the maximum over the 3 nearby bins including the central bin. This would allow the frequency path to deviate by as much as $K-1$ bins. The bin set need not be contiguous or symmetric - this can serve to accommodate large periodic modulation or spindowns.

One issue that arises in practice is that the maximum needs to be taken over power and thus one would need to substitute polarization coefficients into the power sum. When the grid of sampled polarization is not carefully controlled the amount of computation can be large enough to be comparable (or even dominate) the computation of the loosely coherent sum. 

As common to a dynamic programming algorithm the memory requirements for efficient computation can be large as well.

Ideally, the maximum should be computed at the power sum level before substituting polarization coefficients. However, the exact implementation is difficult because the maximum of two quadratic forms or two rational functions is in general a {\em piecewise} rational function. 

This difficulty could be overcome by replacing the piecewise function with a rational function that majorates it - i.e. yields power greater than what would be produced by the exact computation. This gives up some sensitivity for an increase in computational efficiency.

\section{Implementation testing}

\subsection{Execution speed}

Algorithmic improvements tend to have the largest impact on software performance. This is because code optimization usually speeds up a section of the code by at most few hundred percent and the speedup is constant over computation scale. In some situations, such optimization loses with scale as limited hardware features are exhausted when scaling parameter is increased. 

In contrast, an algorithm with smaller power of scaling parameter in asymptotic running time will gain performance with increase in the scaling parameter.

However, sometimes computational hardware is not suitable for a particular algorithm. For example, contemporary CPUs typically do not have dedicated instructions for modular arithmetic, or vector based sorting primitives.

In addition, an actual search has to deal with large numbers of outliers, limited storage and other issues affecting algorithm suitability. 

This makes quantitative comparison of algorithms rather tricky. Ideally, we would want the implementation of each algorithm to be done in the best way possible and the execution conditions to mimic actual search. It seems reasonable to allow optimizing compilers to rearrange code, but, strictly speaking, this changes the algorithm.

To illustrate that the algorithm described in this paper is suitable for contemporary computing hardware we present a comparison with previous loosely coherent implementation. Due to caveats discussed above one should not take this as an actual measure of a speedup in a real search.

Both algorithms were compiled into the same executable using {\tt gcc 6.3.0} compiler and used exactly the same optimization flags. They were both run on O1 1800\,s Hann-windowed SFTs and analyzed the same frequency range 200-200.125\,Hz. The execution node had ample memory and 32 simultaneously executing threads on Intel Xeon E5-2620 processors.

\begin{table*}[htbp]
\begin{center}
\begin{tabular}{lrr}\hline
Parameter & Previous loosely coherent code & New loosely coherent code\\
\hline
\hline
Coherence length (s) & $\sim 14400$  & $14400$ \\
Phase tolerance (rad)  & $\pi/8$ & NA \\
Data load time (s) & 82 & 87 \\
Total running time (s)  & 2579 & 285 \\
Inner loop time (s)  & 2497 & 198 \\
Upper limit  & 4.25e-25 & 3.75e-25 \\
\hline
\end{tabular}
\end{center}
\caption{Execution speed}{Execution speed of loosely coherent algorithms. The new implementation is faster and uses the data more efficiently producing smaller upper limits.}
\label{tab:speed_comparision}
\end{table*}

To save execution time the analysis was restricted to 0.5\,rad disk around Right Ascension 0 and declination 0.

The results are shown in table \ref{tab:speed_comparision}. The previous loosely coherent code refers to the loosely coherent implementation used in \cite{orionspur, O1AllSky1, O1AllSky2}. The column for the new loosely coherent code refers to the implementation of the algorithm described in this paper.

The previous loosely coherent code used phase tolerance as a parameter and a Lanzcos window for the kernel. Because of this, the coherence length is only approximate as the SFTs that took part in the coherent sum were combined with different weights and the window tails covered $43200$ seconds of data.

The coherence length for the new algorithm is better defined, with distance between start of SFTs limited to 14400\,s. Note that because of Hann-windowing the coherent sum made use of data points more than 16000\,s apart but with progressively smaller weights near the ends of the window.

In addition, both codes used real data with duty cycle less than 100\,\%. Thus some coherent sums simply did not have enough data to fill the entire coherence sum.

The upper limit quoted in the table is the worst-case upper limit on all points searched in the grid. The worst-case is typically reached for linear polarizations which introduce strong modulation on SFT weights with approximately 12\,hour period. This further decreases effective coherence length, but the effect is similar for both algorithms.

The data load time describes the time to load all the input time and to perform initialization of all data structures. Total running time describes actual wall-clock time for each code. In the actual application the work load is usually sized to be much larger than input time. 

The inner loop time is the difference between total running time and data load time. It includes the time spent running loosely coherent algorithms as well as the time spent in post-processing their results. The postprocessing algorithm was the same in both instances.

We observe more than 10x improvement in running time, while the upper limit is smaller by 12\,\%. We note that if the phase tolerance parameter was further reduced for the previous loosely coherent code the upper limit would improve, but the running time will greatly increase.

\subsection{Validation}

In this section, we describe validation of the software pipeline implemented using medium scale {\em Loosely coherent} algorithm using Monte-Carlo injections. This demonstrates that the {\em Loosely coherent} algorithm can be implemented on contemporary hardware and it is practical for searches of disks on the sky. Upcoming papers will describe search results.

\begin{table}[htbp]
\begin{center}
\begin{tabular}{rD{.}{.}{2}D{.}{.}{3}}\hline
Stage & \multicolumn{1}{c}{Coherence length (hours)} & \multicolumn{1}{c}{Minimum SNR}\\
\hline
\hline
0  & 8  & 6 \\
1  & 12 & 6.5 \\
2  & 16 & 7 \\
3  & 24 & 8 \\
4  & 36 & 9 \\
5  & 48 & 11 \\
6  & 72 & 13 \\
\hline
\end{tabular}
\end{center}
\caption{Simulation parameters}{Parameters of 6-stage pipeline used for simulations. Stage 6 outliers are subjected to consistency check between interferometers.}
\label{tab:pipeline_parameters}
\end{table}

\begin{figure}[htbp]
  \begin{center}
  \includegraphics[height=8cm]{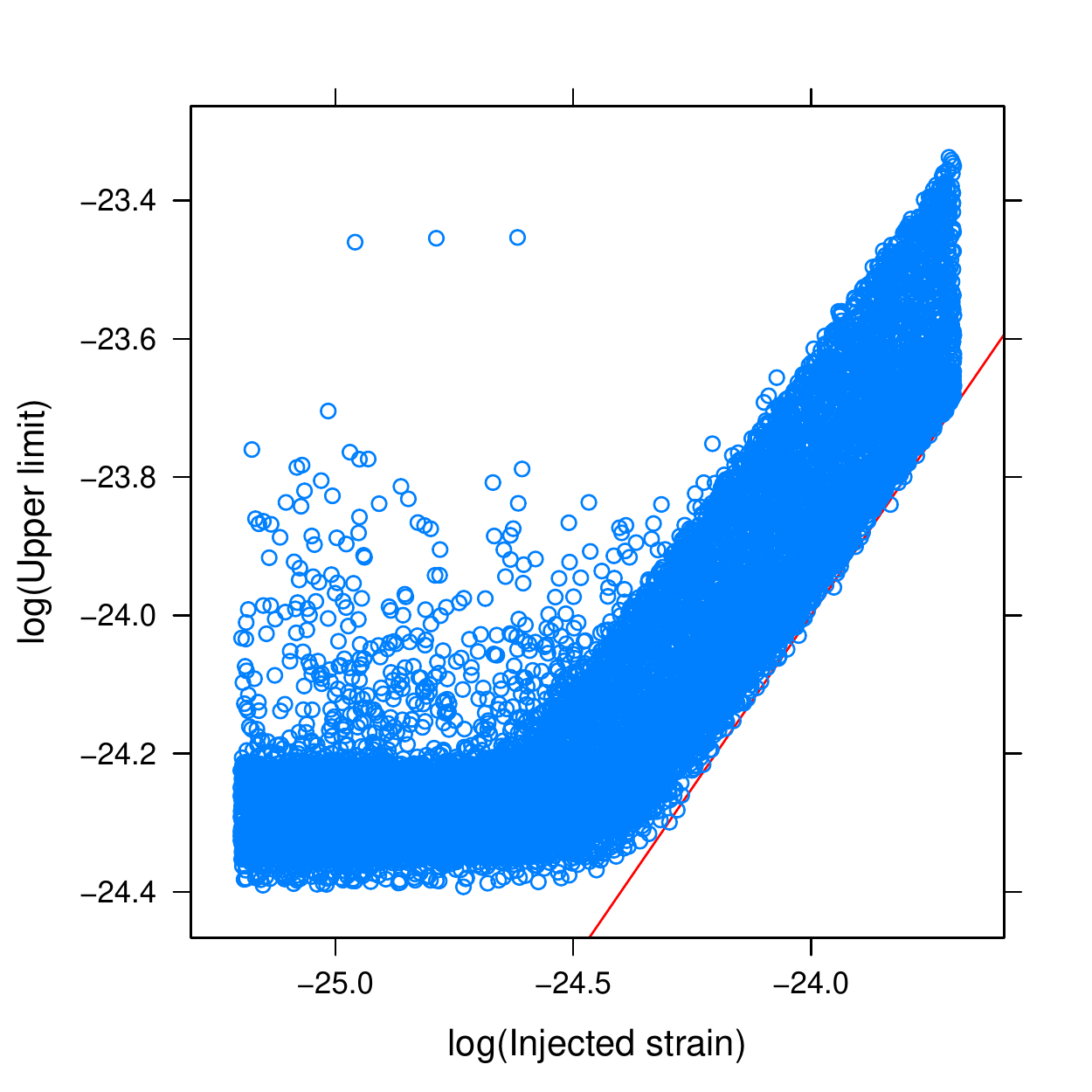}
  \caption{Established upper limit versus injection strength.}
  \label{fig:combined_ul_vs_strain} 
  \end{center}
\end{figure}

%\begin{figure}[htbp]
%  \begin{center}
%  \includegraphics[height=8cm]{combined_ul_vs_strain_rel}
%  \caption{Established upper limit versus injection strength. Both numbers are relative to estimated background level.}
%  \label{fig:combined_ul_vs_strain_rel} 
%  \end{center}
%\end{figure}

\begin{figure}[htbp]
  \begin{center}
  \includegraphics[height=8cm]{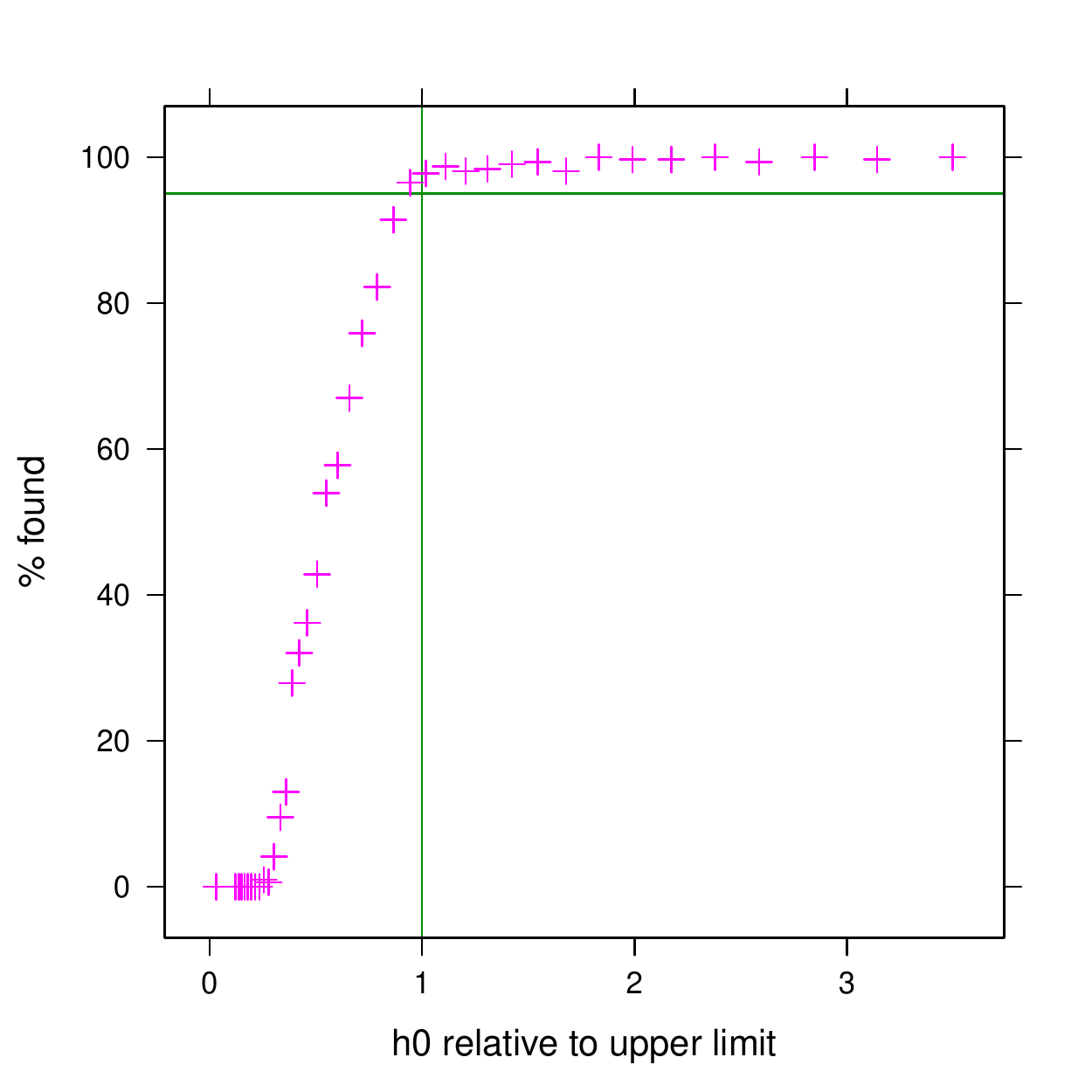}
  \caption{Fraction of injections that had a nearby outlier at the end of the 6-stage pipeline. The X-axis of this figure shows ratio of injection strain to the no-injection upper limit. The Y-axis shows percentage of correctly detected injections. This meant that the injection had to have an outlier nearby that passed all analysis stages. The horizontal green line is drawn at 95\% level.}
  \label{fig:combined_injection_recovery_rel} 
  \end{center}
\end{figure}

The analysis pipeline follows the method described in \cite{O1AllSky1, O1AllSky2}. The practical implementation is a pipeline that progressively increases coherence length (Table \ref{tab:pipeline_parameters}). Each stage, except the last, identifies templates with large upper limits and SNR and writes their parameters to file. These templates are excluded from computation of stage-specific aggregate quantities such as maximum upper limit and SNR. The followup stage analyses templates from the file produced by previous stage.

At the last stage the analysis is done twice - using coherent combination of interferometers and also separately for each interferometer. All outliers are written into the file. The threshold for outlier SNR from individual interferometers is set at 6.  A consistency check is applied, requiring  outliers from coherent combination of interferometers to match outliers from individual interferometers in frequency, sky and spindown.

The pipeline was tested with software injections into O1 data \cite{o1_data} from Advanced LIGO interferometers \cite{aligo}. 
The O1 run occurred between September 12, 2015 and January 19, 2016.

The injections were done randomly into the range of 975-1500\,Hz into a disk on the sky of radius 0.06\,rad ($3.43^\circ$) centered on Right Ascension 4.65\,rad ($266.42^\circ$) and declination $-$0.46\,rad ($-26.35^\circ$). The disk was chosen to cover the center of our galaxy. The spindown of the injections was logarithmically distributed between $-\sci{5.7}{-12}$ and $-\sci{1.8}{-10}$.

The upper limits are computed as a maximum of upper limits from every stage. Figure \ref{fig:combined_ul_vs_strain} shows results of simulation covering range 975-1500\,Hz, where injections were performed into a region near galactic center. The 95\% confidence level upper limits are correctly established, with most points above the red diagonal line. The three outlier points and the scatter of high upper limits for small injection strengths are due to detector artifacts \cite{O1AllSky1, O1AllSky2}. 

%This influence can be accounted for with background estimate. A good proxy for background is the upper limit from first stage of the pipeline obtained with exactly the same parameters but no injection. Figure \ref{fig:combined_ul_vs_strain_rel} shows the same data, but with both injected strain and upper limited divided by a background estimate. 

Figure \ref{fig:combined_injection_recovery_rel} shows detection efficiency of the pipeline. The X-axis of this figure shows ratio of injection strain to the no-injection upper limit. This upper limit  is established in a small sky area (disk with $0.02$\,rad radius) around the injection location by the first stage of the pipeline. In a real search the upper limits are computed as a maximum over larger parameter space, which would shift the X-axis to the right.

The Y-axis of figure \ref{fig:combined_injection_recovery_rel} shows fraction of injections that had a nearby outlier. This outlier had to have an SNR of at least 13, be within $15$\,$\mu$Hz of true injection frequency and within $\sci{1.5}{-11}$\,Hz/s of true injection spindown. 

In addition, nearby outliers had to be no further than $\sci{6.5}{-4} \cdot \left(1\,\textrm{kHz}/f\right)$ in ecliptic distance, defined as the distance between outlier location and true injection location after projection onto the ecliptic plane. Here $f$ is the outlier frequency.

Because high-SNR  outliers are passed to subsequent stages we retain ability to make detection below the upper limit established by the first stage. Thus the effective coherence length of the pipeline is larger than the length defined by the first stage.

\section{Algorithm design and choice of hardware architecture.}

The reader might wonder what would the ideal computer architecture be for the algorithm described in the paper\footnote{We thank the anonymous referee for posing this question.}. Before answering this question it is useful to review contemporary computer architectures.

\subsection{Contemporary and historical computer architecture.}

\begin{figure}[htbp]
  \begin{center}
  \includegraphics[height=8cm]{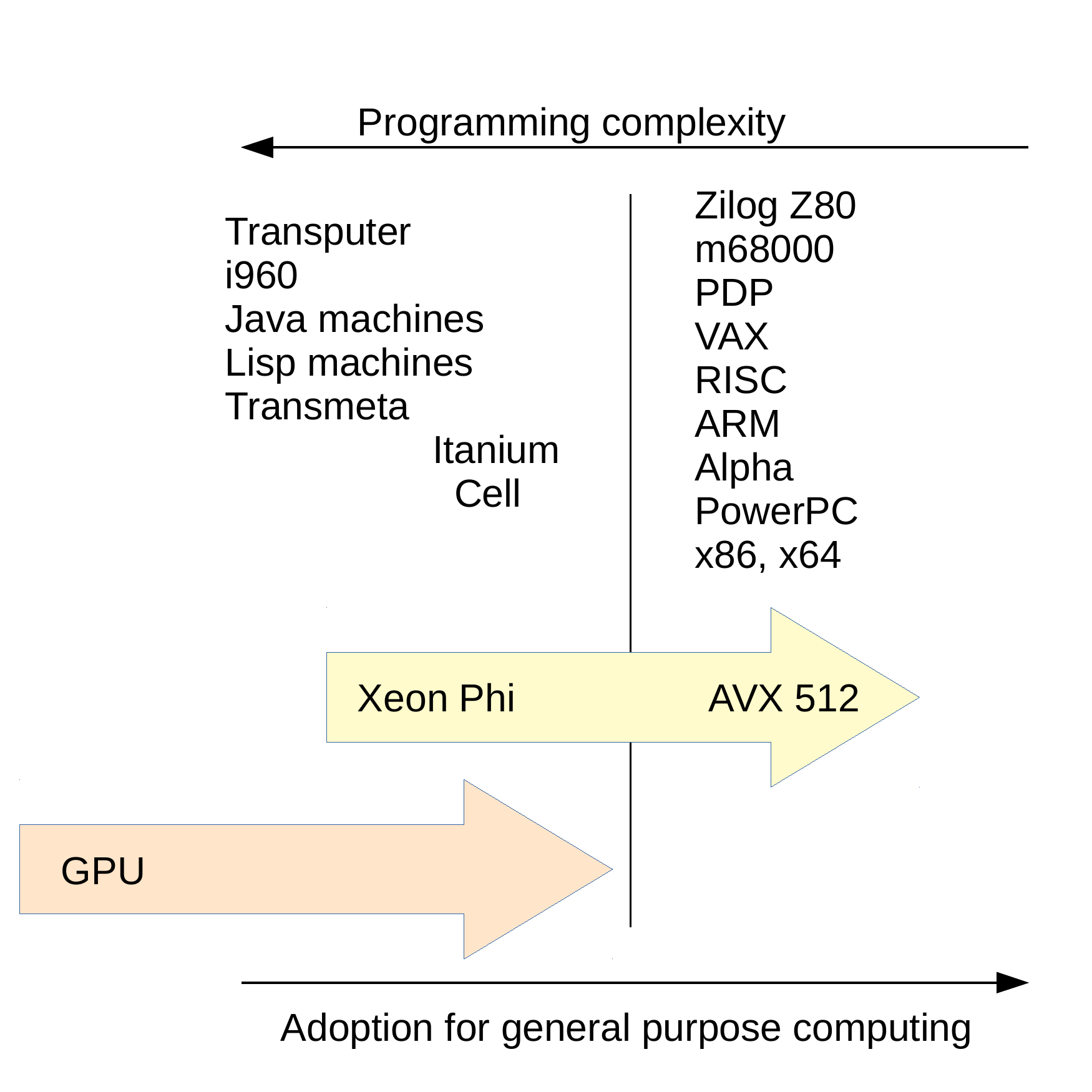}
  \caption{An overview of CPU architectures. We contrast well-known ubiquitous architectures such as Z80 and x64 with exotic systems such as i960 and Cell. In the text we argue that GPUs have more in common with these exotic systems and are slowly changing to look more like AVX512 units.}
  \label{fig:CPU_chart} 
  \end{center}
\end{figure}

Figure \ref{fig:CPU_chart} lists several notable central processing unit (CPU) families. Let us briefly comment on each.

The column on the right lists CPUs that were widely available for extended time:
\begin{itemize}
 \item {\tt Zilog Z80} evolved from an earlier {\tt 8080} CPU by Intel. Implementing 16-bit instructions set, Z80 availability and ease of programming made it popular with both industrial and amateur users. The CPUs are still manufactured today for the embedded market.
  \item {\tt m68000} or {\tt m68k} family was produced by Motorola. This processor was used in early Apple computers and mini-workstations such as Sun or NeXT systems.
  \item {\tt PDP} and {\tt VAX} were very popular series of mini-computers produced by DEC systems. The ease of programming using PDP-11 instructions had large influence on subsequent systems.
  \item {\tt RISC}, {\tt ARM}, {\tt Sparc}, {\tt Alpha} and {\tt PowerPC} were implementation of reduced instruction set computing (RISC) paradigm. They featured small, easy to understand instruction sets that were easy to use and write optimizing compilers for.  
  \item {\tt x86} and {\tt x64} are 32-bit and 64-bit families of CPUs produced by Intel and AMD. Most contemporary computing clusters use these CPUs.
\end{itemize}

The column on the left lists CPUs with innovative approaches to computing that had limited success:
\begin{itemize}
 \item {\tt Transputer} was designed to be used in clusters of identical units. The instruction set was designed to support {\tt Occam} computer language, with later ports of more common languages like {\tt C} or {\tt Fortran}.
  \item {\tt i960} was attempt by Intel to build a CPU with larger and more complicated instructions, but simplified hardware.
  \item {\tt Java machines} were designed to facilitate a safer programming environment. A complimentary Java language is widely used today, with programs running
  in software emulation of Java hardware platform.
  \item {\tt Lisp machines} were designed to run Lisp programs and featured support for many Lisp features, including garbage collection.
  \item {\tt Transmeta} was a company that produced CPUs with very long instructions (VLIW) which were only meant to be used by Transmeta-developed just-in-time compiler to emulate other architectures, in particular x86. These CPUs initially featured much lower power consumption than competitors.
  \item {\tt Itanium} was a 64-bit VLIW CPU produced by Intel. It required a special compiler to properly schedule instructions so that they would not try to use incomplete results of previous calculations. The introduction of much simpler x64 architecture by AMD have superseded Itanium.
  \item {\tt Cell} microprocessor combines a PowerPC core with multiple special purpose cores. The special purpose cores lack branch prediction and cannot access main memory directly, instead relying on a cache-like local memory.
\end{itemize}

From the point of view of algorithm design the computing platforms in the right column are all equivalent - the same code can be compiled for any of them.

The left column on the other hand presents a trap - one can invest a lot of time to take advantage of special hardware, but by the time the code is finished more conventional hardware might catch up. The special purpose hardware might change in the next version requiring extensive code modification, or a complete algorithm redesign. 

Of greater concern are algorithm limitations imposed by the special hardware - you might run faster for the particular computation performed, but you might not be able to use more 
advanced algorithms suitable only for general purpose CPUs.

Right now, there are three main choices in computer clusters: regular CPUs with vector units, graphical processing units (GPUs) and Xeon Phi.

Xeon Phi from Intel was the first commercial product to implement AVX512 instruction set. Similar to Cell it contained rather weak integer CPU cores, coupled with high-performance vector unit that could perform operations on 16 single-precision or 8 double-precision numbers at a time. Unlike Cell, the memory access was unified which greatly simplified programming. Programmatically it acted like a conventional CPU with a top-notch vector unit. Still, making use of full computational performance required careful programming to overcome weakness of in-order integer core and the need for manual prefetching when using non-trivial data layouts. 

As of now GPU-like Xeon Phis have been discontinued, in favor of 
integrating AVX512 vector units into modern CPUs. The compatibility was preserved, so that the code developed for original Xeon Phis is suitable (with a recompile) for new CPUs.

Newer Intel CPUs that implement AVX512 instruction set have more powerful integer cores, making it easier to keep vector units busy and reducing the need for explicit prefetching. Because the vector unit and the integer core share the same cache it is possible to implement algorithms where the vector instructions depend dynamically on computed data. 

Graphical processing units (GPUs) evolved from fixed-function computer graphics units. The original application of 3d graphics required lots of identical operations with little (if any) reuse of computed data. For example, in a 3d visualization applications the CPU would upload a stream of graphical commands to the GPU which would perform computations and store the result in the frame buffer that is displayed on the monitor. When the next frame has completed computing the contents of the frame buffer are not needed anymore and are discarded. 

The GPUs typically lack an equivalent of integer core that can run a {\tt for} loop generating instructions for the vector unit. Instead, a driver program running on the computer the GPU is attached to generates a stream of instruction packets that are stored into the buffer. The instruction packets are then streamed to the GPU via PCI Express bus. 

The driver program can check for completion of computation by reading a single register value. Because these accesses occur over PCI Express bus the latency is hundreds of nanoseconds large. Thus, even if all data used for computation is stored in the GPU-local memory, one incurs a large latency to alter instruction stream based on computed data.

The origin in computer graphics where many independent pixels need to be rendered favoured GPU architecture with large numbers of isolated computational units. Because of this the GPUs are structured to perform many identical operations on data streams that are completely independent of each other. Due to the expectation of little data reuse when rendering real-time graphics, one usually has to  manually manage cache coherence.

The origins in computer graphics also manifest in non-unified address space. For example, NVidia GPUs can access main memory accessible by CPU, texture memory, frame buffer memory, internal memory used for temporary storage, etc. While some GPUs (such as made by AMD) have public documentation describing instruction set and hardware features, others do not (NVidia).

Thus the architecture of contemporary GPUs combines the worst features of i960, Itanium, Cell and Transmeta. Unlike i960\footnote{We note that a notable application of i960 platform was as a graphics processor of NeXT workstations where it run PostScript interpreter that would render directly to the screen.}, GPUs have a firm foothold in graphics accelerator market, which provides resources for further development. 

The evolution of GPUs so far tended to shed features rooted in fixed-function graphics accelerator functionality and add support for wider variety of algorithms and simplified programming. For example, AMD ``Application procession units'' have evolved from merely combining CPU and GPU in the same package to share virtual address spaces and provide cache coherence. They also feature tighter coupling than provided by PCI Express bus.

The experience of Transmeta showed that dynamically generated stream of instructions can 
be more computationally efficient than a static program because it can take advantage of optimizations that are impossible to predict at compilation time.

We would expect therefore that the future GPUs will look more like Xeon Phi/AVX512 units with progressively tighter integration between CPU and GPU core allowing instruction streams to be dynamically generated.

AVX512 will likely add additional instructions implementing reduced precision arithmetic to accommodate artificial intelligence applications (some 16-bit instructions are expected to appear in Ice Lake chips from Intel).

\subsection{Algorithm design strategies}

The overview of computer architectures suggests the following broad guidelines:

\begin{itemize}
 \item Wide floating point vector instructions are here to stay and computationally intensive algorithms should be designed to take advantage of them. The algorithms described in this paper were tested to work well with 8-wide AVX and 16-wide AVX512 instruction sets. We do not see any obstacles to taking advantage of 32-wide instructions.  
 
 \item Successful application of GPU and APU computational hardware requires substantially static floating point instruction streams operating on large numbers of independent inputs. 
 The algorithms described in this paper, with the exception of associative cache, are suitable for this provided a large amount of memory is available to store intermediate results. In such a situation the computation speed is often limited by memory bandwidth. Preliminary tests suggest that realistic analysis requires at least several gigabytes of graphics memory local to the GPU.
 
 \item We expect GPUs and APUs to evolve by acquiring features that bring them closer to conventional CPU architectures. Simultaneously, CPUs will have to compete with computational throughput of GPUs and APUs.  Thus we recommend to use GPUs and APUs only when they provide large computational gains that cannot be offset by algorithm improvement.
\end{itemize}

The algorithms described in this paper have been successfully implemented on general purpose CPUs. The implementation code is structured to take advantage of vector processing units, without substantial increase in code complexity due to the use of vectorizing compilers (both {\tt gcc} and Intel {\tt icc} compiler have provided good results). 

The algorithm itself is suitable for contemporary GPUs with at least several gigabytes of memory, however this comes with substantial burden in code complexity. While periodically we reexamine suitability of GPUs for large scale searches, for now we focus attention on achieving speed improvements through algorithm design first, and wait for GPUs and APUs to evolve to simplify programming and to converge with Xeon Phi-like general purpose CPUs.

In actual usage our implementation needs at least 2\,GB of memory per execution thread. This relatively large memory requirement is due to combination of memory demand by associative caches and dynamic programming code. Thus the ideal hardware would favor faster 
cores with wider vector units. Hyper-threading - presenting more hardware execution threads to increase instruction parallelism - is usually counterproductive to well-tuned codes as it increases memory requirements. The use of associative caches benefits from CPUs with shorter pipelines and smaller penalties for mispredicted branches or memory accesses.

\section{Conclusions}

We have described a new {\em Loosely coherent} algorithm for medium scale coherence lengths. Simulations of a full pipeline incorporating this algorithm have been  demonstrated on O1 data \cite{o1_data, losc} from Advanced LIGO interferometers \cite{aligo}. 

Compared with previous implementation, more than 10x improvement in computational speed has been observed.

The new pipeline allows all-sky loosely coherent analysis for the first time, extending reach of wide-parameter searches for continuous gravitational wave signals. 

\section{Acknowledgments}

The simulations were performed on ATLAS cluster at AEI Hannover. 
We thank Carsten Aulbert and Henning Fehrmann for their support.

The author thanks the LIGO Scientific Collaboration for access to the data and gratefully acknowledge the support of the United States National Science Foundation (NSF) for the construction and operation of the LIGO Laboratory and Advanced LIGO as well as the Science and Technology Facilities Council (STFC) of the United Kingdom, and the Max-Planck-Society (MPS) for support of the construction of Advanced LIGO. Additional support for Advanced LIGO was provided by the Australian Research Council.

This research has made use of data, software and/or web tools obtained from the LIGO Open Science Center (\url{https://losc.ligo.org}), a service of LIGO Laboratory, the LIGO Scientific Collaboration and the Virgo Collaboration. LIGO is funded by the U.S. National Science Foundation. Virgo is funded by the French Centre National de Recherche Scientifique (CNRS), the Italian Istituto Nazionale della Fisica Nucleare (INFN) and the Dutch Nikhef, with contributions by Polish and Hungarian institutes.

We also thank the anonymous referees for many useful comments.

\end{document}